\documentclass{article}
\usepackage{graphicx}
\usepackage{amsmath}

\begin{document}

\title{Understanding Polarization Correlation of Entangled Vector Meson Pairs}

\author{Xun Chen, Siguang Wang\thanks{siguang@pku.edu.cn}, Yajun Mao\thanks{maoyj@pku.edu.cn}\\
  School of Physics and State Key Laboratory of Nuclear Physics and Technology,\\ Peking University, Beijing, 100871, China}
\date{May 30, 2012}

\maketitle

\begin{abstract}
  We propose an experimental test of local hidden variable theories against quantum mechanics by measuring the polarization correlation of entangled vector meson pairs. In our study, the form of the polarization correlation probability is reproduced in a natural way by interpreting the two-body decay of the meson as a measurement of its polarization vector within the framework of quantum mechanics. 
This provides more detailed information on the quantum entanglement, thus a new Monte Carlo method to simulate the quantum correlation is introduced. We discuss the feasibility of carrying out such a test at experiments in operation currently and expect that the measured correlated distribution may provide us with deeper insight into the fundamental question about locality and reality.
\end{abstract}

Debates over the interpretation of quantum mechanics (QM) have lasted for decades. In their famous paper in 1935~\cite{Einstein:1935rr}, Einstein, Podolsky, and Rosen (EPR) argued that QM is incomplete by considering a {\em Gedankenexperiment}, which is now known as the EPR paradox. The paradox challenges the principle of uncertainty that the position and momentum of a particle cannot be precisely measured simultaneously. This paradox also extends to other pair of conjugate physical quantities. In 1951, Bohm expressed the paradox with particle spins in his book~\cite{Bohm:1951qt} and considered a set of alternative theories retaining locality and reality, which are called local hidden variable theories (LHVTs)~\cite{Bohm:1951xw}. It had been considered that one of the LHVTs may replace QM, but Bell concluded that no LHVT can reproduce all the predictions of QM~\cite{Bell:1964kc,Bell:1964fg}, since the predictions of LHVTs would satisfy the Bell's inequality while those of QM may violate it. Other forms of inequalities similar to the original Bell's inequality are derived~\cite{54978,Clauser:1974tg} for the experimental discrimination of the LHVT and QM.

Since 1972, a series of EPR experiments have been carried out using the entangled photon pairs~\cite{Freedman:1972zz, Aspect:1981zz, Aspect:1981nv, Ou:1988zz, Shih:1988zz}, and their results favored QM. But conclusions can not be drawn decisively because these experiments could not be treated loophole free~\cite{Santos:1992zz}.
Tests of the Bell's inequality in particle physics have also been considered.
As far back as the 1960s, the entanglement of $K^0\bar{K}^0$ pair was noticed and studied~\cite{Lipkin:1969nd}.
Violation of the Bell's inequality have been observed in the $B^0\bar{B}^0$ mesons from $\Upsilon(4S)$ decay by the Belle collaboration~\cite{Go:2003tx, Go:2007ww}. However it was argued that the inequality used is not a genuine Bell's inequality so that it can't make a discrimination between QM and LHVTs~\cite{Bramon:2004pp}. 
T\"ornqvist suggested that the weakly decay particles, such as the $\Lambda\bar{\Lambda}$ pairs generated from the decay of $J/\psi$ or $\eta_c$, could be used for testing of nonlocality of QM, assuming the CP invariance~\cite{Tornqvist:1980af}. It has been studied by the DM2 collaboration with about $10^3$ $J/\psi\to\Lambda\bar{\Lambda}$ events~\cite{Tixier:1988fv}, but the statistics is not sufficient to give a decisive conclusion.
Test of local realism with the vector meson pairs from the pseudoscalar cascade decays, such as $\eta_c\to V V\to (PP)(PP)$, was also suggested in recent years~\cite{Li:2008dk, arXiv:0903.1246}.

In this article, we try to provide a straightforward understanding of the entanglement between the vector mesons within the framework of QM. We will show that the deduction leads to a new Monte Carlo (MC) simulation method for the correlated decay, and the understanding implies that local realism could be tested in high energy physics (HEP) experiments by measuring the correlated distribution.

\begin{figure}
  \centering
  \includegraphics[width=8.5cm]{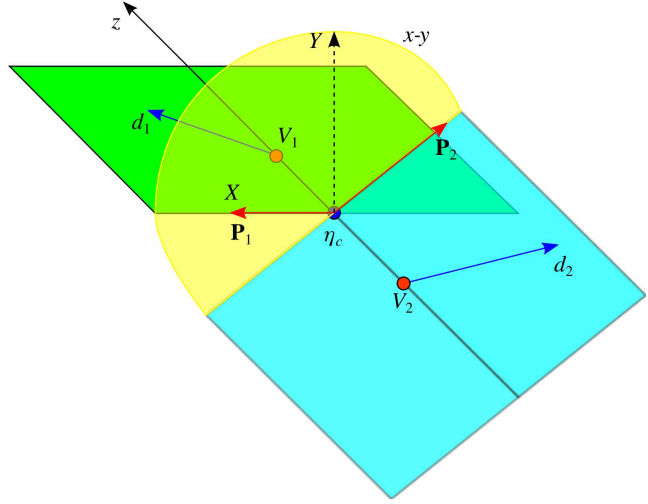}
  \caption{Decay of a pair of entangled vector mesons $V_1V_2$ from $\eta_c$'s decay. $\mathbf{P}_1$ and $\mathbf{P}_2$ are the polarization vector of meson $V_1$ and $V_2$, respectively.}
  \label{fig:decaydis}
\end{figure}

We start from a detailed description of the entangled state of the vector meson pair from the decay of $\eta_c$.
A sketch of the cascade decay process of $\eta_c$ to vector meson pair $V_1V_2$ is illustrated in Fig.\ \ref{fig:decaydis}. The $z$-axis is chosen to be parallel to meson $V_1$'s momentum.
Due to $CP$ invariance in strong interaction, the total spin of the vector meson pair is $S=1$. The conservation of angular momentum requires the total spin projection on the $z$-axis is zero. Then the two mesons are transversely polarized and the wave function of the system can be written as~\cite{Li:2008dk,arXiv:0903.1246}
\begin{equation}
  \label{eq:entanglewave_z}
    |\Psi\rangle =  \displaystyle\frac{1}{\sqrt{2}}\left(|1\rangle_{z}|\mbox{-}1\rangle_{z} - |\mbox{-}1\rangle_{z}|1\rangle_{z} \right),
\end{equation}
where the subscript $z$ means the $z$-axis.
It can also be expressed in terms of bases in the $x$-$y$ plane as
\begin{equation}
  \label{eq:entanglewave_alpha}
   |\Psi\rangle  =  \displaystyle\frac{1}{\sqrt{2}}\left(|0\rangle_{\alpha_\perp}|0\rangle_{\alpha} - |0\rangle_{\alpha}|0\rangle_{\alpha_\perp}\right),
\end{equation}
where the subscript $\alpha$ means an arbitrary axis in the $x$-$y$ plane, and $\alpha_\perp$ indicates another axis on the same plane and perpendicular to $\alpha$. The wave functions $|0\rangle_{\alpha}$ and $|0\rangle_{\alpha_\perp}$ are the eigenfunctions of spin operators $J_{\alpha}$ and $J_{\alpha_\perp}$, respectively. This form means that when we measure the spin of one meson at the $\varphi$ direction and get zero value, the spin of another meson at $\varphi_\perp$ can be deduced to be zero too. It should be noted that Eq.~\ref{eq:entanglewave_z} and Eq.~\ref{eq:entanglewave_alpha} are physically equivalent, and the state of the system does not change by varying the form of wave function.

It has been pointed out that the decay of the mesons can be regarded as a ``measurement'' of their polarization vectors~\cite{arXiv:0903.1246}.
The direction of the measurement is given by the momentum projection of the daughter particles on the $x$-$y$ plane.
For example, when the meson $V_1$ decays into $d_1$ and another particle, its polarization $\mathbf{P}_1$ is measured on the same direction as that of the momentum projection of $d_1$ on the $x$-$y$ plane(see Fig.~\ref{fig:decaydis}).
Another daughter particle of $V_1$ is not illustrated as it has the opposite momentum component on the $x$-$y$ plane to that of $d_1$. The meson $V_2$'s polarization $\mathbf{P}_2$ is measured in the same way.

After the decay of both mesons, the polarization vectors $\mathbf{P}_1$ and $\mathbf{P}_2$ do not need to be orthogonal, as the ``measurement'' leads to quantum decoherence.
When one meson decays, its polarization vector $\mathbf{P}_1$ is measured at an azimuthal angle $\alpha$. The entanglement requires the polarization vector of the second meson to have an azimuthal angle of $\alpha\pm\pi/2$ according to Eq.\ (\ref{eq:entanglewave_alpha}). That is similar to the case that when we measure the spin of one of the entangled spin-$1/2$ particles in one direction, another one should have a definite spin value in the same direction.
However, the polarization vector of the second meson is finally determined by a measurement, i.e., its own decay. The decay of first meson changes the probabilities of the measurement of the polarization vector of the second meson.

QM and LHVTs give different predictions on the probability of $\mathbf{P}_1$ with azimuthal angle $\alpha_1$ and $\mathbf{P}_2$ with azimuthal angle $\alpha_2$. The prediction of QM violates one form of Clauser-Horne-Shimony-Holt inequality~\cite{arXiv:0903.1246}. In QM, the probability can be obtained directly with the wave function in Eq.~(\ref{eq:entanglewave_alpha}).

The correlated probability can also be calculated in another way with above interpretation of ``{\em decay as measurement}.'' We write the correlated probability as $P(\alpha_1, \alpha_2)$, where $\alpha_{1,2}$ are the azimuthal angles of the polarization vectors $\mathbf{P}_1, \mathbf{P}_2$ respectively. Due to the rotation invariance, the probability depends only on the difference $\Delta\alpha=\alpha_2-\alpha_1$. So we can write the probability as $P(\Delta\alpha)$.

In Fig.~\ref{fig:decaydis}, the axis with the same direction of $\mathbf{P}_1$ is marked as $X$, which is the $\alpha$-axis in Eq.~(\ref{eq:entanglewave_alpha}), and $Y$ is the $\alpha_\perp$-axis. The probability $P(\Delta \alpha)$ is the same as the conditional probability that $\mathbf{P}_2$ taking an azimuthal angle $\Delta \alpha$ given that the azimuthal angle of $X$ is fixed at the azimuthal angle 0.

\begin{figure}
  \centering
  \includegraphics[width=8.5cm]{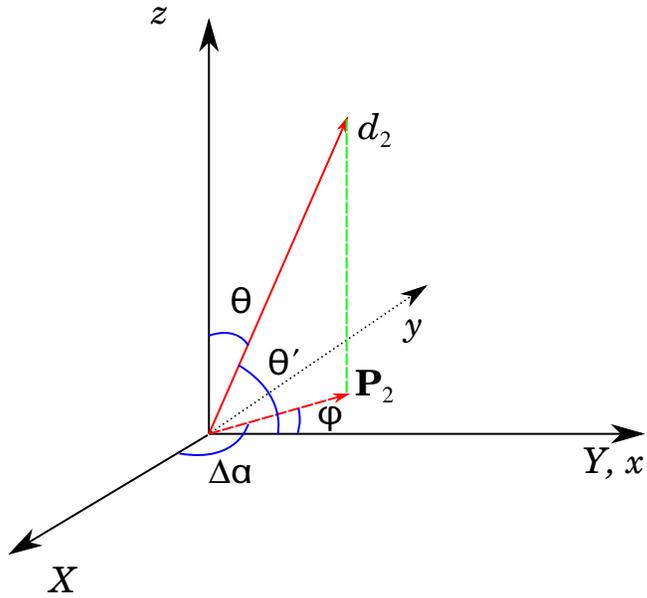}
  \caption{A schematic drawing of the coordinate system and angles. The $X$ and $Y$-axes have the same meaning as that given in Fig.~\ref{fig:decaydis}. The $xyz$ coordinate system can be obtained by rotating the original $XYz$ around the $z$-axis for $\pi/2$. The angles $\theta$ and $\varphi$ angles are defined in the $xyz$ coordinate system. The angle $\theta'$ is defined as the angle between $d_2$ and the $Y$-axis.}
  \label{fig:coordinate}
\end{figure}

Because the decay acts as a measurement, the angular distribution of final state particles of meson $V_2$ determines the probability $P(\Delta\alpha)$. Without loss of generality, if the angular distribution of $V_2$'s daughter particle $d_2$ is described by the probability $P(\cos\theta, \varphi)$ in the $xyz$ coordinate system given in Fig.~\ref{fig:coordinate}, we have
\begin{equation}
  \label{eq:correlated_dis}
  \begin{array}{rl}
    P(\Delta\alpha) = & \int_{-1}^{1}P(\cos\theta, \varphi)d\cos\theta\cdot \left| \displaystyle\frac{\partial \varphi}{\partial\Delta\alpha}\right | \\
    = & \int_{-1}^{1}P(\cos\theta, \varphi)d\cos\theta,
  \end{array}
\end{equation}
where $\varphi = \Delta\alpha-\pi/2$. 

Within the framework of QM, the angular distribution of the decay of a single particle can be described by the helicity formalism~\cite{CALT-68-1148}, which is widely used in particle physics. 
For the two-body decay, as the two final state particles are back-to-back in their center-of-mass frame, their directions can be characterized by the polar angle $\theta$ and azimuthal angle $\varphi$. The cross section of a spin-1 vector meson decay into two spin-0 particles can be written down as
\begin{equation}
  \label{eq:kdis_1}
  \begin{array}{rcl}
    \displaystyle\frac{d\sigma(\theta, \varphi)}{d\Omega} = \frac{d\sigma}{d\cos\theta d\varphi}  & = & \displaystyle\frac{3}{4\pi} |D^1_{M0}(\varphi, \theta, -\varphi)A|^2 \\
    & \propto & |Y_1^M(\theta, \varphi)|^2,
  \end{array}
\end{equation}
where $D^1_{M0}$ is the Wigner D-Matrix $D^j_{m\lambda}$ with spin $j=1$, and spin-projection $m=M$ and $\lambda=0$, and $A$ is a matrix element without any angular dependence. $Y_1^M(\theta, \varphi)$ is the spherical harmonic function with $l=1$. $M$ depends on the selection of spin quantization axis.

The decay of meson $V_1$ fixes its polarization vector on the $X$-axis, then the entanglement between $V_1$ and $V_2$ requires that the polarization vector of $V_2$ is on the $Y$-axis before a measurement is applied on it. That is, for the decay of $V_2$, the spin quantization axis should be chosen to be the $Y$-axis, with $m=0$, so the cross section has an angular dependence of $|Y_1^0(\theta', \varphi')|^2\propto \cos^2\theta'$, where the polar angle $\theta'$ is the angle between the daughter particle's momentum and the $Y$-axis(see Fig.~\ref{fig:coordinate}), and the azimuthal angle $\varphi'$ is defined in the $X$-$z$ plane.

Eq.~(\ref{eq:correlated_dis}) requires us to write the angular distribution with $\theta$ and $\varphi$ in the original $xyz$ coordinate frame. It can be obtained via a simple transformation:
\begin{equation}
  \label{eq:h10rote}
  \begin{array}{rl}
    P(\cos\theta, \varphi) = & P(\cos\theta', \varphi') \left|\displaystyle \frac{\partial (\cos\theta', \varphi')}{\partial (\cos\theta, \varphi)} \right| \\
    = & P(\cos\theta', \varphi') \\
    \propto & \cos^2\theta',
  \end{array}
\end{equation}
where the Jacobian determinant is equal to 1 due to the rotation invariance.

With the definition of coordinate system in Fig.~\ref{fig:coordinate}, the relations between angles $(\theta, \varphi)$ and $(\theta',\varphi')$ can be written down:
\begin{equation}
  \label{eq:transform}
  \begin{array}{rcr}
    \cos\varphi'\sin\theta' & = & -\cos\theta,\\
    \sin\varphi'\sin\theta' & = & \sin\varphi\sin\theta, \\
    \cos\theta'& = & \cos\varphi\sin\theta.
  \end{array}
\end{equation}
Thus, we obtain
\begin{equation}
  \label{eq:y10}
  P(\cos\theta, \varphi) \propto \sin^2\theta\cos^2\varphi.
\end{equation}
Substituting above formula into Eq.~(\ref{eq:correlated_dis}), we arrive at
\begin{equation}
  \label{eq:pdis_final}
  P(\Delta \alpha) \propto \cos^2\varphi = \sin^2\Delta \alpha.
\end{equation}
The polarization correlation can be calculated directly with QM.
It is given by~\cite{Li:2008dk},
\begin{equation}
  \label{eq:correlation}
    P(\alpha_1, \alpha_2) = \displaystyle\frac{1}{2}\sin^2\Delta\alpha.
\end{equation}
Eqs.~(\ref{eq:pdis_final}) and (\ref{eq:correlation}) give the same form of polarization correlation. That means our understanding of the entanglement is compatible with QM.

One important application of the understanding is that it provides a new method to simulate the quantum entanglement in the process of cascade decay of $\eta_c\to VV\to (PP)(PP)$ directly. Sampling the directions of final state particles is one of the essential tasks in MC simulation.
From our deduction, it becomes very straightforward. At first, either of the two vector mesons $V_1$ is selected. Its decay can be treated as if it is not affected by other particles, thus we can set it quantization axis to be the $z$-axis with $m=1$. Its daughter particles would follow the angular distribution of $|Y^{\pm1}_1(\theta, \varphi)|^2 \propto \sin^2\theta$.
Once the angles $\theta_1$ and $\varphi_1$ of $V_1$'s daughter particles are sampled, the quantization axis of the second meson $V_2$ must be chosen on the $x$-$y$ plane and is fixed at the direction of $\varphi\pm\pi/2$. One should use the angular distribution of $|Y^{0}_1(\theta', \varphi')|^2\propto \cos^2\theta'$ to sample the angles $\theta_2'$ and $\varphi_2'$. To obtain the angles $\theta_2$ and $\varphi_2$, a rotation is performed for an unit vector constructed with $(\theta_2',\varphi_2')$, just as that given in Fig.~\ref{fig:coordinate}.

\begin{figure}
  \centering
  \includegraphics[width=8.5cm]{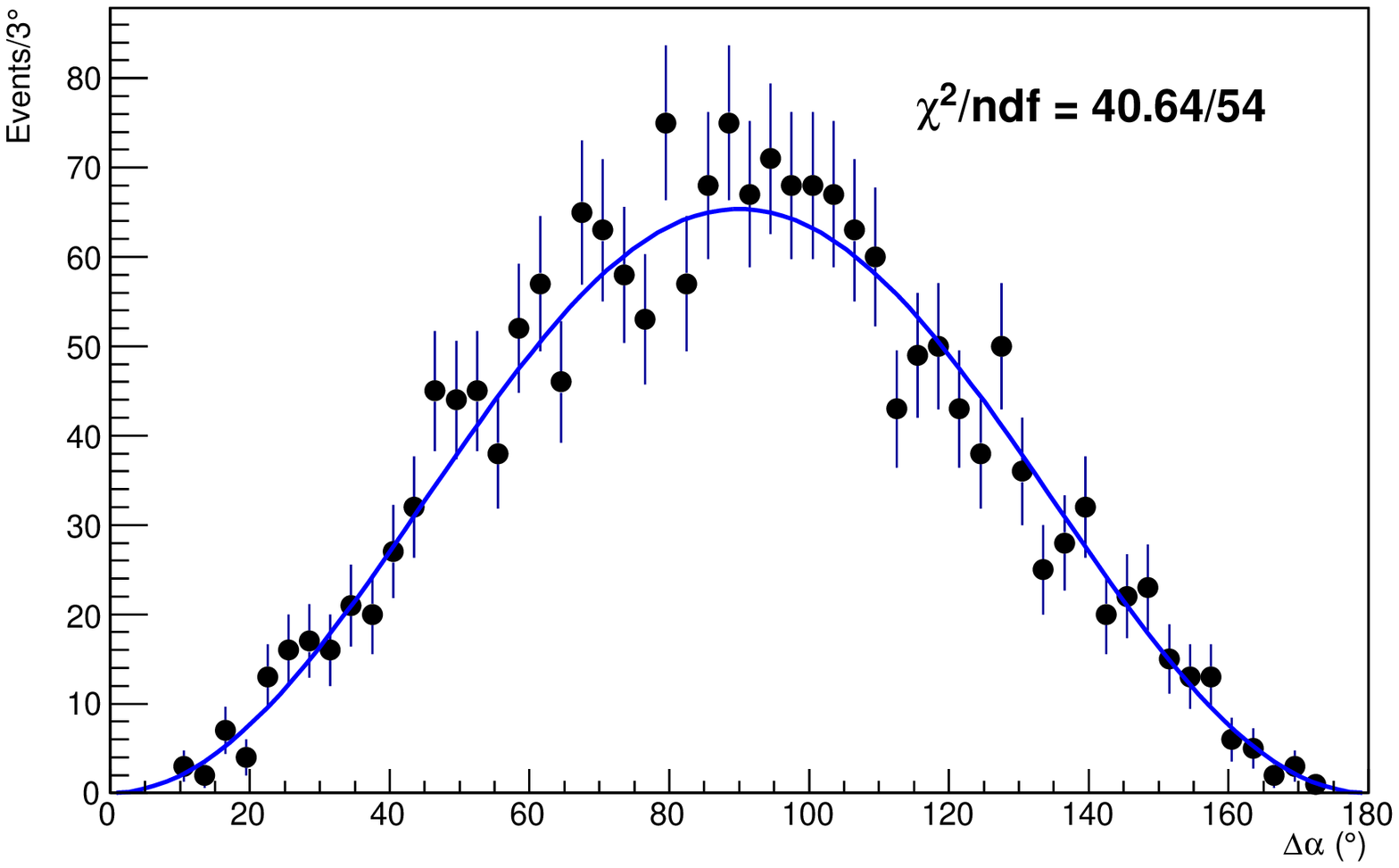}  
  \caption{The correlated distribution of the polarization vectors of the two vector mesons, 2,000 events. The $x$-axis is the angle between the polarization vectors. The distribution follows the formula of $P(\Delta \alpha)\propto\sin^2\Delta\alpha$.}
  \label{fig:correlation}
\end{figure}

An implementation of our MC method for the process of $\eta_c\to\phi\phi\to K^+K^-K^+K^-$ is done.
To check whether the simulation method works as expected, 2,000 events are generated, and the correlated distribution $P(\Delta\alpha)$ is extracted. The result is plotted in Fig.\ \ref{fig:correlation}.
The distribution is fitted with the function of $p\cdot\sin^2\Delta\alpha$. It is clear that the distribution satisfies those given in Eq.\ (\ref{eq:pdis_final}) and (\ref{eq:correlation}). The method works well.

More importantly, the understanding suggests a way to discriminate QM and LHVTs in HEP experiments by measuring the correlated distribution of polarization.
A variation of the Clauser-Horne-Shimony-Holt inequality has been derived in Refs.\ ~\cite{Li:2008dk, arXiv:0903.1246} based on the correlated probability $P(\alpha_1, \alpha_2)$. By substituting the form of $P(\alpha_1, \alpha_2)$ of QM into the inequality, one can verify that the inequality is violated at some selected values of $\alpha_1, \alpha_2$. In other words, if the observed correlated distribution follows the prediction of QM, the inequality {\em must} be violated, thus LHVTs are ruled out.

A realistic chance to perform such a test is using the cascade decay of $J/\psi \to \gamma\eta_c\to\gamma\phi\phi\to\gamma(K^+K^-)(K^+K^-)$ in a $J/\psi$-factory such as the Beijing Spectrometer III (BES-III) experiment, in which $10^9$ $J/\psi$ particles could be produced in several months~\cite{majp:2009}. The branching fraction of the cascade decay $J/\psi\to \gamma\eta_c \to \gamma\phi\phi \to \gamma(K^+K^-)(K^+K^-)$ is about $1.07\times 10^{-5}$~\cite{Nakamura:2010zzi}. By assuming the detector efficiency of $20\%$ at the BES-III experiment, one would expect that about $2,000$ desired events can be collected for analysis.
MC simulation shows that when the number of collected data events is larger than 1,000, the correlated distribution could be extracted from data with acceptable level of confidence.

One disadvantage of the HEP experiments is the existence of backgrounds, which may distort the final results. For example, the processes of $J/\psi \to \gamma \phi\phi\to\gamma (K^+K^-)(K^+K^-)$, $\eta_c\to \phi K^+K^-\to (K^+K^-)(K^+K^-)$ and $\eta_c\to 2(K^+K^-)$ give the same final state particles as those of our desired process. Fortunately, the last two process have small branching fractions, and can be suppressed to the order of around $1\sim2\%$ of signals by limiting the masses of reconstructed $\phi$ and $\eta_c$ mesons~\cite{Wang:2012an}. The processes from $J/\psi$ to $\gamma\phi\phi$ are the main sources of backgrounds, but they are not studied well. A former study shows the number of background events is about $40\%$ of that of the signal events within the mass region of $\eta_c$, but its statistics is low~\cite{Bai:1990hk}. The BES-III experiment is expected to have lower background level with improved detectors and higher statistics. At current stage, we could assume that there is no correlation within the background events for simplicity. This assumption leads to the uniform distribution of $P(\Delta \alpha)$ for backgrounds. The final distribution would take the form of $\sin^2\Delta\alpha +c$, where $c$ is a constant, and the fitted number of background events could be used to eliminate it. In general, if the distribution shape of the background is known, one can subtract the background from the total distribution, and thus suppress the impact of the background on the final result.

In the traditional ways to test local realism, especially in the optical experiments, some selected data points at which the Bell's inequalities are violated are measured. The disadvantage of those methods is that they might rule out the LHVTs, but couldn't tell whether QM is valid, because it is difficult to obtain the predicted correlated distribution.
The advantage of our method is noticeable. High statistics in HEP experiments make it possible to obtain the distribution of $\Delta\alpha$, so that it may provide a further answer for the fundamental question on which theory to favor. Though the existence of background may distort the signal, modern data analysis technologies could be employed to correct the distortion from backgrounds and other sources.

In summary, we reproduce the same form of correlated distribution between entangled mesons predicted by QM by interpreting the two-body decay of a meson as the measurement of its polarization vector. This provides more information on the detail of quantum entanglement.
Thus, a new MC method to simulate the quantum correlation in the pseudoscalar cascade decay is introduced. 
With this understanding, we propose that the measurement of correlated distribution in HEP experiments can be used to make a discrimination between QM and LHVT. This is feasible in existing experiments, such as the BES-III experiment. The measured correlated distribution would provide a further answer for the fundamental question on which theory, QM or LHVT, is favored by nature.

\section*{Acknowledgment}
The authors thank Professor Cong-Feng~Qiao, Dr.~Junli~Li, Dr.~Peng~Sun and Dr.~Jian-Hui~Zhang for the valuable discussions. The authors also would like to thank Mingsu~Cong, Zijun~Xu, Zhongming~Qu, Jiangbo~Wei for the related former research. This work is supported by the Postdoctoral foundation of Peking University, China Postdoctoral Science Foundation (2012M510249) and the National Natural Science Foundation of China (10979010, 10835001).

\begin{thebibliography}{99}

\bibitem{Einstein:1935rr} 
  A.~Einstein, B.~Podolsky and N.~Rosen,
  Phys.\ Rev.\  {\bf 47}, 777 (1935).

\bibitem{Bohm:1951qt}
  D.~Bohm,
  {\it Quantum Theory}, Prentice-Hall, Englewood Cliffs, (1951).

\bibitem{Bohm:1951xw} 
  D.~Bohm,
  Phys.\ Rev.\  {\bf 85}, 166 (1952).

\bibitem{Bell:1964kc} 
  J.~S.~Bell,
  Physics {\bf 1}, 195 (1965).

\bibitem{Bell:1964fg} 
  J.~S.~Bell,
  Rev.\ Mod.\ Phys.\  {\bf 38}, 447 (1966).

\bibitem{54978} 
  J.~F.~Clauser, M.~A.~Horne, A.~Shimony and R.~A.~Holt,
  Phys.\ Rev.\ Lett.\ \ {\bf 23}, 880  (1969).

\bibitem{Clauser:1974tg} 
  J.~F.~Clauser and M.~A.~Horne,
  Phys.\ Rev.\ D {\bf 10}, 526 (1974).

\bibitem{Freedman:1972zz} 
  S.~J.~Freedman and J.~F.~Clauser,
  Phys.\ Rev.\ Lett.\  {\bf 28}, 938 (1972).

\bibitem{Aspect:1981zz} 
  A.~Aspect, P.~Grangier and G.~Roger,  
  Phys.\ Rev.\ Lett.\  {\bf 47}, 460 (1981).

\bibitem{Aspect:1981nv} 
  A.~Aspect, P.~Grangier and G.~Roger,
  Phys.\ Rev.\ Lett.\  {\bf 49}, 91 (1982).

\bibitem{Ou:1988zz} 
  Z.~Y.~Ou and L.~Mandel,
  Phys.\ Rev.\ Lett.\  {\bf 61}, 50 (1988).

\bibitem{Shih:1988zz} 
  Y.~H.~Shih and C.~O.~Alley,
  Phys.\ Rev.\ Lett.\  {\bf 61}, 2921 (1988).

\bibitem{Santos:1992zz} 
  E.~Santos,
  Phys.\ Rev.\ A {\bf 46}, 3646 (1992).

\bibitem{Lipkin:1969nd} 
  H.~J.~Lipkin,
  Phys.\ Rev.\  {\bf 176}, 1715 (1968).

\bibitem{Go:2003tx} 
  A.~Go  [Belle Collaboration],
  J.\ Mod.\ Opt. {\bf 51}, 991 (2004)
  quant-ph/0310192.

\bibitem{Go:2007ww} 
  A.~Go {\it et al.}  [Belle Collaboration],
  Phys.\ Rev.\ Lett.\  {\bf 99}, 131802 (2007).

\bibitem{Bramon:2004pp} 
  A.~Bramon, R.~Escribano and G.~Garbarino,
  J.\ Mod.\ Opt.\  {\bf 52}, 1681 (2005).

\bibitem{Tornqvist:1980af} 
  N.~A.~T\"{o}rnqvist,
  Found.\ Phys.\  {\bf 11}, 171 (1981).

\bibitem{Tixier:1988fv} 
  M.~H.~Tixier {\it et al.}  [DM2 Collaboration],
  Phys.\ Lett.\ B {\bf 212}, 523 (1988).


\bibitem{Li:2008dk} 
  J.~Li and C.~-F.~Qiao,
  Phys.\ Lett.\ A\ {\bf 373}, 4311 (2009).


\bibitem{arXiv:0903.1246}J.~Li and C.~-F.~Qiao,
  Sci.\ China G\ {\bf 53}, 870  (2010).

\bibitem{CALT-68-1148} 
  J.~D.~Richman,
  DOE Research and Development Report
  No. CALT-68-1148, 1984

\bibitem{majp:2009} 
  J.-P.~Ma,
  Int.\ J.\ Mod.\ Phys.\ A {\bf 24}, 3 (2009).

\bibitem{Nakamura:2010zzi} 
  K.~Nakamura {\it et al.}  [Particle Data Group Collaboration],
  J.\ Phys.\ G {\bf 37}, 075021 (2010).

\bibitem{Wang:2012an}
  M.~Cong, Z.~Xu, S.~Wang, Z.~Qu, J.~Wei,
  Nucl.\ Instrm.\ Meth.\ A {\bf 679}, 14 (2012).

\bibitem{Bai:1990hk} 
  Z.~Bai {\it et al.}  [MARK-III Collaboration],
  Phys.\ Rev.\ Lett.\  {\bf 65}, 1309 (1990).

\end{thebibliography}
\end{document}